\documentclass[aps,prc,twocolumn,nofootinbib,showpacs,preprintnumbers,amsmath,amssymb]{revtex4}

\usepackage{graphicx}
\usepackage{dcolumn}
\usepackage{bm}
\usepackage{epsfig}
\usepackage{longtable} 
\def\la{\mathrel{\mathpalette\fun <}}
\def\ga{\mathrel{\mathpalette\fun >}}
\def\fun#1#2{\lower3.6pt\vbox{\baselineskip0pt\lineskip.9pt
\ialign{$\mathsurround=0pt#1\hfil##\hfil$\crcr#2\crcr\sim\crcr}}}

\begin{document}

\preprint{}

\title{
Precise measurement of $\alpha_K$ for the 65.7-keV $M$4 transition in $^{119}$Sn: Extended test of internal-conversion theory
}

\author{N. Nica}
\email{nica@comp.tamu.edu}

\author{J.C. Hardy}
\email{hardy@comp.tamu.edu}

\author{V.E. Iacob}

\author{M. Bencomo}

\author{V. Horvat}

\author{H.I. Park}

\author{M. Maguire}
\altaffiliation {REU summer student from Reed College, Portland, OR}

\author{S. Miller}
\altaffiliation {REU summer student from Florida A\&M University, Tallahassee, FL}

\affiliation{ Cyclotron Institute, Texas A\&M University, College Station, Texas 77843, USA}
\homepage{http://cyclotron.tamu.edu/}

\author{M.B. Trzhaskovskaya}
\affiliation{Petersburg Nuclear Physics Institute, Gatchina 188300, Russia}

\date{\today}

\begin{abstract}
We have measured the $K$-shell internal conversion coefficient, $\alpha_K$, for the 65.7-keV
$M$4 transition in $^{119}$Sn to be 1621(25).  This result agrees well with Dirac-Fock calculations in which
the effect of the $K$-shell atomic vacancy is accounted for, and disagrees with calculations in which the
vacancy is ignored.  This extends our precision tests of theory to $Z$ = 50, the lowest $Z$ yet measured.
 
\end{abstract}


\maketitle

\section{\label{sec:introd} INTRODUCTION}

Internal conversion makes a critical contribution to the majority of nuclear decay schemes.  Yet its contribution
is usually not measured but rather is obtained from tabulated internal conversion coefficients (ICCs), which are
applied to the measured relative gamma-ray intensities to establish total transition intensities.  Decay schemes
built this way are thus highly dependent on the reliability of tabulated ICCs.  The same is true if a measured ICC
is to be used to determine a transition multipolarity or mixing ratio.

In spite of their importance, until recently the accuracy of calculated ICCs was, at best, ill defined.  Very few
precise measurements of ICCs existed at all, and the collected body of less precise results appeared, for many years, 
to be systematically displaced from calculations by a few percent.  In 1973 Raman {\it et al.}~\cite{Ra73} compared
``precisely measured" ICCs for fifteen $E$3 and $M$4 transitions with the tabulated Hager and Seltzer calculations
\cite{Ha68} and concluded that the theoretical values were systematically higher by 2-3\%.  However, even this select
group of transitions included only five with measured ICCs that were known with a precision of 2\% or better, so the
apparent discrepancy was hardly definitive.  Nevertheless, this is where the matter remained for 30 years.

By 2002, Raman {\it et al.}~\cite{Ra02} had 100 experimental ICCs to compare with tabulated values,
but even at that recent date only 20 of the measured ICCs had a relative precision of 2\% or better.  Their
results still indicated that all previous tables of ICCs exhibited a 3\% systematic bias, but the
authors found much better agreement (within $\sim$1\%) for a then-new table by Band {\it et al.}~\cite{Ba02},
which had been calculated in the framework of the Dirac-Fock method, with the exchange between electrons
treated exactly.  However, there was a price to be paid.  The best agreement with data was achieved with a version
of this calculation that ignored the atomic vacancy created by the conversion process: The final-state
electron wave function was computed in a field that did not include any provision for a vacancy.  Whatever
its apparent benefit, this was a patently unphysical assumption since $K$-vacancy lifetimes are known \cite{Ke74}
to be longer than the time for the conversion electron to leave the vicinity of the atom.

The question of whether or not to include the atomic vacancy was settled in favor of its inclusion by our precise
2004 measurement ($\pm$0.8\%) of the $K$-shell conversion coefficient, $\alpha_K$, for the 80.2-keV $M4$ transition
in $^{193}$Ir \cite{Ni04, Ni05}.  The ICCs for this transition calculated with and without the atomic vacancy differed from one another
by more than 10\%.  Our result agreed with the physically reasonable calculation that included the vacancy, and differed by more
than 10 standard deviations from the no-vacancy calculation.  While this appeared quite definitive, we had only tested
a single transition at a single value of $Z$, so we set about making further tests over a wider range of atomic numbers.
Up till now, we have confirmed the need to include the vacancy by measurements of $\alpha_K$ for the 127.5-keV $E3$ transition
in $^{134}$Cs, and the 661.7-keV $M4$ transition in $^{137}$Ba \cite{Ni07, Ni08}.  We also measured the 346.5-keV $M$4 transition in
$^{197}$Pt \cite{Ni09}, which corrected an old result that disagreed with both types of calculation.

By 2008, our early results from this program influenced a reevaluation of ICCs by Kib\'{e}di $et~al.$ \cite{Ki08} who also
developed BrIcc, a new data-base obtained from the basic code by Band $et~al.$ \cite{Ba02} but, in conformity with our
conclusions, it employed a version that incorporated the ``frozen orbital" approximation to account for the atomic hole.  The
BrIcc data-base has been adopted by the National Nuclear Data Center (NNDC) and is available on-line for the determination of ICCs.
Our experimental results obtained since 2008, already referred to, continue to support that decision and have extended our
verification tests over the range $55<Z<78$.   

We report here a measurement that extends that range down to $Z$ = 50.  We have measured the $\alpha_K$ value for
the 65.7-keV $M$4 transition in $^{119}$Sn to a precision of $\pm$1.5\%.  This is quite sufficient precision to distinguish between
the two models for calculating the $\alpha_K$ -- one with, and the other without the atomic vacancy -- which differ from one another by 4.8\%.

\section{\label{sec:overview} Measurement Overview}

We have described our measurement techniques in detail in previous publications \cite{Ni04,Ni07}
so only a summary will be given here.  If a decay scheme is dominated by a single transition
that can convert in the atomic $K$ shell, and a spectrum of $K$ x rays and $\gamma$ rays is recorded
for its decay, then the $K$-shell internal conversion coefficient for that transition is given by
\begin{equation}
\alpha_K \omega_K = \frac{N_K}{N_\gamma} \cdot \frac{\epsilon_\gamma}{\epsilon_K},
\label{alpha}
\end{equation}
where $\omega_K$ is the fluorescence yield; $N_K$ and $N_{\gamma}$ are the total numbers of observed
$K$ x rays and $\gamma$ rays, respectively; and $\epsilon_K$ and $\epsilon_\gamma$ are the
corresponding photopeak detection efficiencies.

The fluorescence yield for tin has been measured several times,
with a weighted average good to $\pm$1.2\% \cite{Hu94}.  Furthermore, world data for fluorescence yields
have also been evaluated systematically as a function of $Z$ \cite{Sc96} for all elements with $10 \leq Z
\leq 100$, and $\omega_K$ values have been recommended for each element in this range.  The recommended value
for tin, $Z$ = 50, is 0.860(4), which is consistent with the average measured value but has a smaller relative
uncertainty, $\pm$0.5\%.  We use this value.

\begin{figure}[t]
\epsfig{file=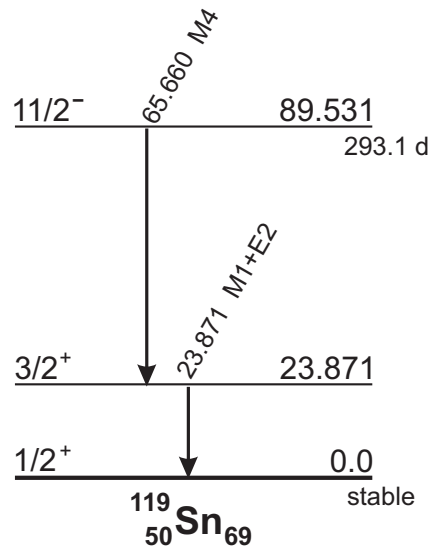,width=5.5cm}
\caption{Decay scheme for the 293.1-day isomer in $^{119}$Sn.  The data are taken from Ref.\,\cite{Sy09}.}
\label{fig1}
\end{figure}

The decay scheme of the 293.1-day isomer in $^{119}$Sn is shown in Fig.~\ref{fig1}.  It has a unique
decay path of two cascaded transitions with energies of 65.7 and 23.9 keV, the former being
the $M4$ transition of interest here.  Both transitions convert but the 23.9-keV transition can only
convert in the $L$ and higher shells.  Thus the $K$ x-ray peak from tin observed in a decay
spectrum of $^{119m}$Sn can only be due to the conversion of the 65.7-keV transition.  This satisfies
the ``single transition" requirement for the validity of Eq.~\ref{alpha}.

In our experiments, we detect both the $\gamma$ ray and the $K$ x rays in the same HPGe detector, a detector
whose efficiency has been meticulously calibrated \cite{Ha02,He03,He04} to sub-percent precision, originally
over an energy range from 50 to 3500 keV but more recently extended \cite{Ni08} down to 34 keV, the average
energy of lanthanum $K$ x rays.  Over this whole energy region, precise measured data were combined with Monte Carlo
calculations from the CYLTRAN code to yield a very precise and accurate detector efficiency curve.  In our present study of the $M$4
transition in $^{119}$Sn, the $\gamma$ ray of interest is at 65.7 keV, which is well within the calibrated region, but
the tin $K$ x rays lie between 25 and 29 keV, somewhat below even our extended region of calibration.  We will describe
in Sec.~\ref{sec:Cal} a measurement of the decay of $^{109}$Cd, which we use to extend our range of calibration even
further, down to 22 keV, the energy of the $K_{\alpha}$ x ray of silver.  Our energy range of calibration thus
encompasses the energies of both the $\gamma$ ray and the $K$ x rays from the decay of $^{119m}$Sn.  However its
precision at the lowest energies cannot match what we have achieved above 50 keV, so the efficiency ratio,
$\epsilon_{\gamma}/\epsilon_K$, we will use in Eq.~\ref{alpha} for this case has a relative precision of $\pm$1\%.

\section{\label{sec:experiment} Experiment}

We used the same experimental method and setup as in our previous measurements
\cite{Ni04,Ni05,Ni07,Ni08,Ni09}.  Only those details not covered in previous publications will be
described here.

\subsection{\label{sec:sourceprep} Source preparation}

We produced $^{119m}$Sn by neutron activation of a tin metal foil enriched to 98.8\% in $^{118}$Sn.
The foil had been rolled to a thickness of 6.8 $\mu$m by the supplier, Trace Sciences International
Corporation.  The company specified that chemical impurities in the material totaled 10 ppm at most,
and that the only significant isotopic impurities were $^{117}$Sn (0.8\%) and $^{119}$Sn (0.4\%),
with the sum of all other stable isotopes of tin contributing $<$0.1\%.  We cut the foil into two 1-cm squares.

For two reasons we anticipated the need for a very long activation time: First, the cross section for
producing $^{119m}$Sn from thermal-neutron activation of $^{118}$Sn was known to be small \cite{Mu81},
0.010(6)\,b; and second, the total ICC for the 65.7-keV transition is $\sim$5000 \cite{Sy09}, which means
that the 65.7-keV $\gamma$ ray is very weak compared to the tin x rays and requires a strong source to
generate sufficient statistics for a precise measurement in a reasonable amount of time.  Unfortunately,
these conditions can also result in impurity activities dominating the decay spectrum and potentially
affecting the peaks of interest. 

As an initial test we exposed one foil for 12 hours to a thermal neutron flux of $\sim7\times10^{12}$ n/cm$^2$s
at the TRIGA reactor in the Texas A\&M Nuclear Science Center.  After removal from the reactor the
sample was conveyed to our measurement location, where counting began a week after the end of
activation. The initial activity from $^{119m}$Sn was determined to be 2.2 kBq, predominantly seen as
x rays of course.  Although far from negligible, the impurity activities were deemed to be manageable.

The second foil was then exposed to the same neutron flux for 120 hours.  At the end of this time, the
foil had acquired a yellowish hue and had adhered to the aluminum can in which it had been contained.
However, with the help of a razor blade, we successfully removed a large portion of the foil, which was still
flat enough to use without any increase in self attenuation of the tin x rays.  The initial activity of
the $^{119m}$Sn contained in this recovered foil was 22 kBq, but the total activity of the foil was very much
higher than that.  Consequently, we did not record a spectrum for analysis until twelve weeks after activation, 
by which time the short-lived impurity activities had died away.

\subsection{\label{sec:Rad decay} Radioactive decay measurements}

We acquired spectra with our precisely calibrated HPGe detector and with the same electronics used in
its calibration \cite{He03}.  Our analog-to-digital converter was an Ortec TRUMP$^{TM}$-8k/2k card
controlled by MAESTRO$^{TM}$ software.  The TRUMP$^{TM}$ card uses the Gedcke-Hale method \cite{Je81}
to determine a live time that accounts for dead-time losses and random summing.  We acquired 8k-channel
spectra at a source-to-detector distance of 151~mm, the distance at which our calibration is well established.
Each spectrum covered the energy interval 10-2000 keV with a dispersion of about 0.25 keV/channel. 

Our first spectrum, begun three months after activation, was recorded for a total of nearly two months.  It was
carefully analyzed for impurities by the methods described in Sec.~\ref{sec:Imp ident}.  Two impurities in
particular were discovered to be of serious concern: $^{75}$Se ($t_{1/2}$ = 119.8\,d) and $^{182}$Ta
($t_{1/2}$ = 114.4\,d).  Both produce $\gamma$ rays that overlapped the 65.7-keV $\gamma$ ray from the
$^{119m}$Sn decay transition of interest, creating a single broad group at 66 keV.  In spite of valiant efforts
to correct for these impurities, it became clear to us that the final result for
$\alpha_K$ would not be of sufficient precision to distinguish between the two ICC calculations we were
investigating.

Fortunately, the half-lives of $^{75}$Se and $^{182}$Ta are less than half that of $^{119m}$Sn, so time
alone solved our problem.  We waited more than a year and then recorded our second spectrum, just under two
years after the end of activation.  We took it under the same conditions as the first spectrum, and
recorded it for a total of 24 days.  It was this spectrum alone, with room background subtracted, that we used
in the analysis we report here.

Even after this prolonged delay, a cluster of $^{182}$Ta peaks, both $\gamma$ rays and x rays, remained strong
enough to interfere with the $^{119}$Sn 65.7-keV $\gamma$ ray.  To obtain a template for these interfering peaks we
also activated a 32.8-$\mu$m-thick tantalum foil (99.988\% $^{181}$Ta in natural abundance) for 10 s in
the TRIGA reactor, and subsequently recorded the $^{182}$Ta decay spectrum for 19 days. 

We mentioned in Sec.~\ref{sec:overview} that we also measured the decay of $^{109}$Cd in order to extend
the energy range of our detector's efficiency calibration.  This was done with a 160-kBq source covered by
0.25-mm-thick aluminized Mylar, purchased from Eckert and Ziegler Isotope Products.  It was placed in exactly
the same geometry as was used for our measurement of $^{119m}$Sn.

We made one further auxiliary measurement.  As can be deduced from Fig.~\ref{fig1}, in addition to the peaks of
interest, the decay of $^{119m}$Sn also produces a 23.9-keV $\gamma$ ray.  This peak is only partially resolved
from the $K_{\alpha}$ x-ray peak, which appears at 25.2 keV.  In order to be fully confident of our extraction of the
x-ray peak area, we measured the $^{119m}$Sn spectrum with a 6-mm-diameter, 5.5-mm-deep Si detector, which we placed
102 mm from the source.  This detector has much higher resolution than the germanium detector (but has too low an
efficiency at 65.7 keV to be suitable for the complete ICC measurement).  The 23.9-keV peak was cleanly separated
from the $K$ x rays in the Si-detector spectrum, allowing us to determine its relative intensity.

\section{\label{sec:analysis} Analysis}

In our analysis of the $^{119m}$Sn data, as well as the $^{109}$Cd calibration data, we followed the same
methodology as we did with previous source measurements \cite{Ni04,Ni05,Ni07,Ni08, Ni09}.  We first extracted
background-subtracted areas for essentially all the x- and $\gamma$-ray peaks in the spectrum.  Our procedure was
to determine the areas with GF2, the least-squares peak-fitting program in the RADWARE series \cite{Rapc}.  In
doing so, we used the same fitting procedures as were used in the original detector-efficiency calibration
\cite{Ha02,He03,He04}.

Once the areas (and energies) of peaks had been established, we could identify all impurities in the $^{119m}$Sn
spectrum and carefully check to see which were known to produce x or $\gamma$ rays that interfered with the
tin $K$ x rays or the 65.7-keV $\gamma$ ray, our peaks of interest.  Then we took the spectrum we measured
from the commercial $^{109}$Cd source and extracted areas for its x-ray peaks and the 88.0-keV $\gamma$-ray peak
so as to extend our detector's energy region of well-calibrated efficiency down to 22 keV.  With the full range of
efficiencies known, we could return to the $^{119m}$Sn spectrum and make appropriate corrections to account for the
effects of observed impurities on the peaks of interest.  Next, with the help of the Si-detector spectrum, we
refined our determination of the $K$ x-ray peak area.  Finally we dealt with the various small corrections that
had to be applied to the areas of the x-ray peaks to take account of self attenuation and of their non-Gaussian shape. 

\begin{figure}[t]
\epsfig{file=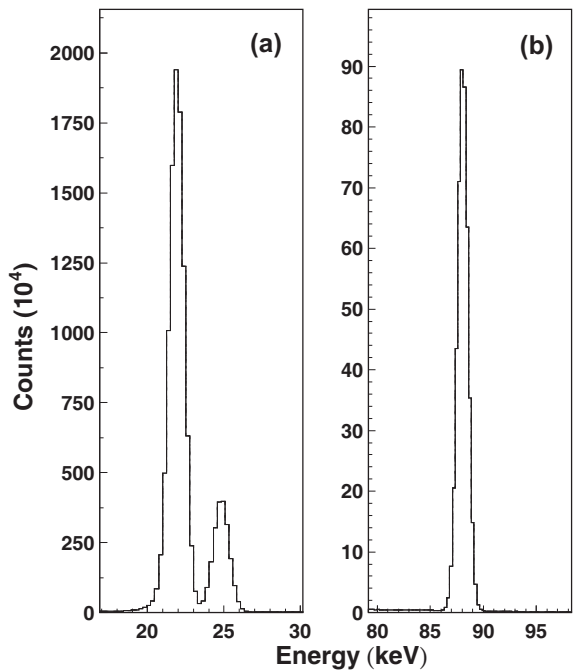,width=7.5cm}
\caption{Portions of the energy spectrum recorded for a $^{109}$Cd source.  (a) The left-hand panel shows the silver $K_{\alpha}$ and
$K_{\beta}$ x-ray peaks at 22 and 25 keV respectively.  (b) The right-hand panel shows the single 88-keV $\gamma$-ray peak.}
\label{fig2}
\end{figure}

\subsection{\label{sec:Cal} Low-energy efficiency calibration}

The two energy regions of interest in the $^{109}$Cd decay spectrum are shown in Fig.~\ref{fig2}, where it is evident
that the peak areas could easily be determined with precision.  In using the result to establish detector efficiency, however,
we need to know the emission probability for $K$ x-rays relative to that for 88-keV $\gamma$ rays.  The $^{109}$Cd
decay begins with a pure electron-capture transition that uniquely populates the 88-keV, first-excited state in $^{109}$Ag.
That state then decays by an $E$3 transition which converts.  Both processes, electron capture and internal conversion, cause
$K$-shell vacancies, with subsequent emission of silver $K$ x-rays.

Rewriting Eq.~\ref{alpha} to include the contribution from electron capture, and rearranging it so as to yield the detector
efficiency for the 88.0-keV $\gamma$ rays relative to the efficiency for the $K$ x rays , we obtain
\begin{equation}
\frac{\epsilon_\gamma}{\epsilon_K} = \frac{\omega_K (\alpha_K P_{\gamma} + P_{ec,K})}{P_\gamma} \cdot \frac{N_\gamma}{N_K},
\label{effK}
\end{equation}
where the new parameters, $P_{\gamma}$ and $P_{ec,K}$, are the probabilities per parent decay for $\gamma$-ray emission and
electron capture respectively.  We took $P_{\gamma}$ = 0.03626(20) from the careful analysis published by the IAEA \cite{IA07},
and $P_{ec,K}$ = 0.8131(10) from a calculation with the LOGFT code available from the National Nuclear Data Center (NNDC) web
site \cite{NNDC2}.  The fluorescence yield, $\omega_K$, for silver is 0.831(4) \cite{Sc96}

The calculated $\alpha_K$ value for the 88-keV transition also depends on whether the atomic vacancy is accounted for or not but, in this case,
the two values only differ by less than 3\%: 11.10 (no vacancy) and 11.41 (vacancy included in the ``frozen orbital" approximation).
So as not to prejudice our ultimate result for $^{119m}$Sn, we adopt the value 11.25$\pm0.16$, which encompasses both values.
Substituting these values into Eq.~\ref{effK}, we obtain the result
\begin{equation}
\frac{\epsilon_\gamma}{\epsilon_K} = 27.98(22)\cdot \frac{N_\gamma}{N_K}.
\label{eff109Cd}
\end{equation}
Note that we have not distinguished between the $K_{\alpha}$ and $K_{\beta}$ x rays. Since scattering effects are difficult to
account for individually when the two peaks are so close together and, in the case of $^{119m}$Sn, impurity contributions are difficult to
distribute between the two peaks, we have chosen throughout this work -- for both $^{109}$Ag and $^{119m}$Sn -- to deal only with the
sum of their $K_{\alpha}$ and $K_{\beta}$ peaks.  For calibration purposes, we consider each sum to be located at the intensity-weighted
average energy of the two component peaks -- 22.57 keV for silver and 25.77 keV for tin.

\begin{table}[b]
\caption{\label{table1} Quantities used in applying Eq.~\ref{eff109Cd} to determine the detector efficiency at 22.57 keV based on our
measurement of the $^{109}$Cd source}
\vspace{2mm}
\begin{ruledtabular}
\begin{tabular}{lll}
Quantity & ~~~~~~~Value &  Uncertainty as \%      \\
\hline  \\[-2mm]
~~~$N_K$  & 1.14577(11)$\times10^8$ & 0.009  \\
~~~$N_{\gamma}$ & 4.378(3)$\times10^6$ & 0.06  \\
~~~$\epsilon_{\gamma\,88.0}$/$\epsilon_{K22.6}$  & 1.069(8)  & 0.8   \\[2mm]
~~~$\epsilon_{\gamma\,88.0}$ \cite{He03} & 1.0030(15)\% & 0.15   \\
~~~$\epsilon_{K22.6}$ & 0.938(8)\% & 0.8   \\
\end{tabular}
\end{ruledtabular}
\end{table} 

\begin{figure*}[t]
\epsfig{file=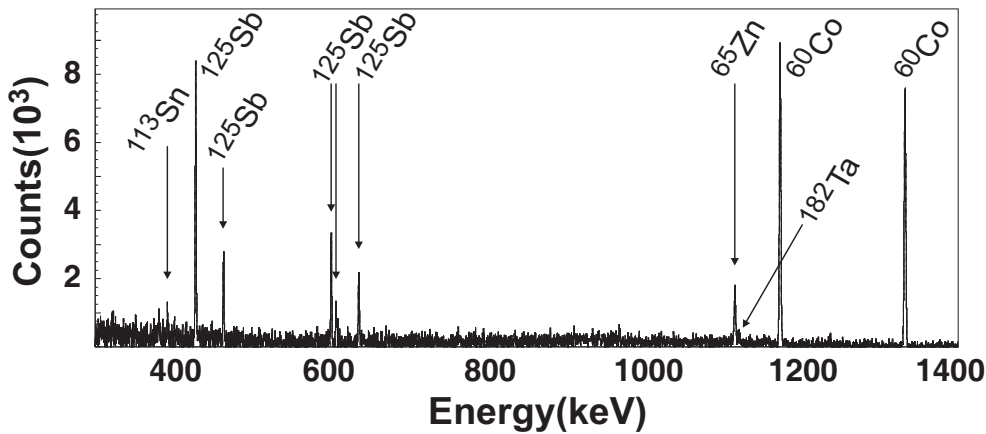,width=13cm}
\caption{Portion of the $\gamma$-ray energy spectrum measured two years after activation of the $^{118}$Sn foil.  Peaks are
labeled by their $\beta$-decay parent.}
\label{fig3}
\end{figure*}

The experimental results for $^{109}$Cd are presented in Table~\ref{table1}.  It should be noted that the value of $\epsilon_K$ derived
there includes the contribution of photons emitted from the source that are Compton-scattered from materials (including air) in the
neighborhood and subsequently recorded by the detector either within the full-energy peak or just to its left.  We have investigated this effect in
detail in a previous measurement \cite{Ni07} and determined it to be of order 0.8\% for x rays close to this energy region.  Here though,
we will be comparing the efficiencies for silver and tin x rays, which lie very close to one another in energy.  Since the scattering contributions
to both should be very nearly the same, we eliminate one source of uncertainty by simply including scattered photons in the peak areas, and in
the detector efficiencies used to analyze them.

Based on the value determined for $\epsilon_{K22.6}$ in Table~\ref{table1}, we made a very small change in one geometrical input parameter in our CYLTRAN
Monte Carlo calibration code: We increased the thickness of our HPGe detector's front dead layer from 2.5 $\mu$m \cite{He03} to 4.25 $\mu$m.  This has
no effect on efficiencies above about 60 keV but extends our calibration down to 22.6 keV.

\subsection{\label{sec:Imp ident} Impurities}

Because of the very low neutron-activation cross section to produce $^{119m}$Sn, even weak impurities in the original
$^{118}$Sn sample become a serious concern if their activation cross sections are relatively large.  Indeed, as mentioned
in Sec.~\ref{sec:Rad decay}, the spectrum we took starting three months after activation was overwhelmed with such impurities,
so much so that we could not use it for a precise $\alpha_K$ determination.  However it did provide us an opportunity to
identify impurities from $\gamma$ rays with statistically significant intensities.  This knowledge proved useful in our
ultimate analysis of the spectrum recorded nearly two years later, in which the impurities were much less intense but, in
some cases, still significant enough to demand attention. 

As a typical example, a portion of our later spectrum appears in Fig.\,\ref{fig3} with contaminant peaks identified by their
parent isotope.  As is evident from the figure, even the weakest peaks were identified.  In all, in the spectrum recorded two
years after activation, we identified seven long-lived contaminant activities. Two affected the tin x-ray energy region, $^{113}$Sn
and $^{125}$Sb; another two, $^{75}$Se and $^{182}$Ta, produced $\gamma$ rays very near our 65.7-keV $\gamma$ ray; and three more,
$^{60}$Co, $^{65}$Zn and $^{133}$Ba, had no impact on our results.

\begin{table}[b]
\caption{\label{table2} The contributions of identified impurities to the energy regions of the 65.7-keV peak and
the tin $K$ x-ray peaks.}
\vspace{2mm}
\begin{ruledtabular}
\begin{tabular}{lll}
Source & Contaminant &  Contaminant       \\
& & contribution (\%)  \\
\hline  \\[-2mm]
\multicolumn{3}{l}{Contribution to $K$ x-ray peaks:}  \\[1mm]
~~~$^{113}$Sn & In $K$ x rays & 0.0154(5)  \\
~~~$^{125}$Sb & Te $K$ x rays & 0.815(18)  \\[2mm]
\multicolumn{3}{l}{Contribution to 65.7-keV peak:}  \\[1mm]
~~~$^{75}$Se & 66.1-keV $\gamma$ ray & 2.46(5)   \\
~~~$^{182}$Ta & 65.7-keV $\gamma$ ray & 4.61(8)   \\
\end{tabular}
\end{ruledtabular}
\end{table}

In the case of $^{113}$Sn, the 392-keV peak identified in Fig.~\ref{fig3} is the strongest peak in its well known
electron-capture decay \cite{Bl10}.  From the known intensity of this $\gamma$ ray in the decay scheme, the number of counts
we observe in the peak, and the established efficiency of our detector, we could then determine the total activity of the
$^{113}$Sn in our sample.  Since the intensity of indium $K$ x rays emitted in this decay is also well known \cite{Bl10},
it was a simple matter to determine the contribution of these x rays to our tin x ray peaks, which are not resolved from them.
The result is given in Table~\ref{table2}.  It is seen to be extremely small.

There are four $\gamma$-ray peaks visible in Fig.~\ref{fig3}, which are attributed to the $\beta$ decay of $^{125}$Sb. The peak
at 428 keV is in fact the strongest in its decay spectrum \cite{Ka11}.  As we did with $^{113}$Sn, we used the observed
$\gamma$-ray intensities to determine the total activity of $^{125}$Sb.  Then, taking the known intensity of tellurium $K$ x rays
emitted in this decay \cite{NNDC1}, we established their contribution to our $K$ x-ray energy region.  The result also appears in
Table~\ref{table2}.

A similar procedure was used to account for the two impurities that affected the energy region around 65.7 keV.  Although
not included in the partial spectrum shown in Fig.~\ref{fig3}, $^{75}$Se was identified and quantified via several $\gamma$ rays
below 300 keV, of which the strongest is at 265 keV.  From the measured peak areas and the known decay scheme of $^{75}$Se
\cite{Fa99}, we derived the intensity of the 66.1-keV $\gamma$ ray.  Its relative contribution to the 65.7-keV peak of interest
from $^{119m}$Sn decay is given in Table~\ref{table2}.

\begin{figure}[t]
\epsfig{file=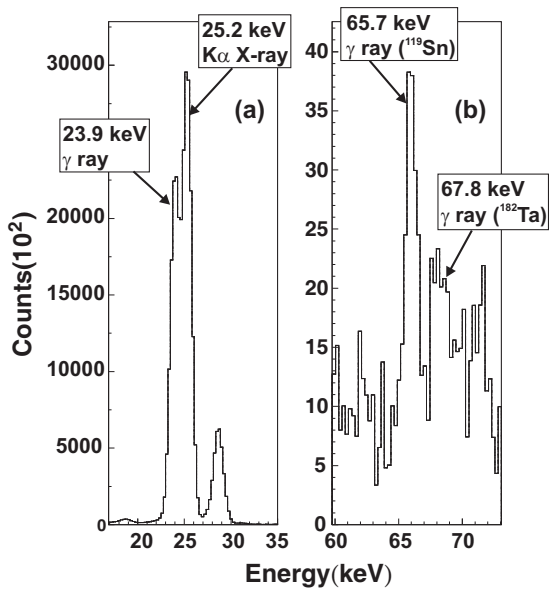,width=7.2cm}
\caption{Portions of the spectrum collected two years after activation, covering the two energy regions of interest
in the determination of the $\alpha_K$ ratio for the 65.7-keV transition in $^{119}$Sn.  (a) The left panel shows,
and identifies, the 23.9-keV $\gamma$-ray peak from the decay of the first excited state in $^{119}$Sn, which is
barely resolved from the tin $K_{\alpha}$ x-ray peak; the third peak in the spectrum is from the tin $K_{\beta}$
x-rays.  (b) The right panel shows the 65.7-keV $\gamma$-ray peak from $^{119m}$Sn, and the nearby impurity peak from
the decay of $^{182}$Ta.  In fact, there are several $^{182}$Ta peaks in this energy region; the 67.8-keV $\gamma$
ray is simply the most prominent.} 
\label{fig4}
\end{figure}

The decay of $^{182}$Ta offered quite a number of peaks as well, including the one identified in Fig.~\ref{fig3}.  This decay is
equally well documented \cite{Si10,NNDC1}.  However, its contribution to the energy region around 65.7 keV is more complex.
Figure~\ref{fig4} shows portions of the measured spectrum centered on the $K$ x rays of tin and on the 65.7-keV $\gamma$-ray peak.
Identified in the right-hand panel is a 67.8-keV peak, one of the strongest from the decay of $^{182}$Ta.  It is easily separable
from the $^{119}$Sn $\gamma$ ray at 65.7 keV.  However, $^{182}$Ta also has a $\gamma$-ray peak at 65.7 keV: It is weaker -- only
7\% the intensity of the 67.8-keV peak -- but it must nonetheless be accounted for.  In addition there are three relatively weak
$K_{\beta}$ x-ray peaks from $^{182}$Ta decay between 67 and 69 keV.  We used a spectrum measured from the decay of a separately
prepared $^{182}$Ta source (see Sec.\,\ref{sec:Rad decay}) as a template to determine the contribution of this contaminant to the
65.7-keV peak from $^{119m}$Sn. The result is given in Table~\ref{table2}.  It turns out that $^{182}$Ta is responsible for the
largest contaminant contribution to our peaks of interest but, even so, it proves to have an inconsequential effect on
the total uncertainty attached to the $\alpha_K$ value for the 65.7 keV transition from $^{119m}$Sn.

\subsection{\label{xray} $^{119m}$Sn x-ray peaks}

\begin{figure}[b]
\epsfig{file=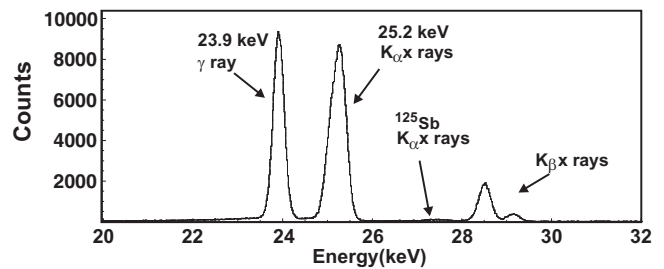,width=8.5cm}
\caption{Spectrum obtained from our $^{119m}$Sn source viewed with a Si detector.  The energy range covered is similar to that
covered in the left panel of Fig.~\ref{fig4}, but with a dispersion of approximately 20 eV per channel.} 
\label{fig5}
\end{figure}

From the left panel of Fig.~\ref{fig4} it can be seen that the 23.9-keV $\gamma$ ray from the decay of $^{119m}$Sn cannot be
well resolved by our HPGe detector from the $K_{\alpha}$ x rays emitted in the same decay.  Yet, its area must be accurately
separated if we are to use the $K$ x-ray peaks to obtain $\alpha_K$ for the 65.7-keV transition.  To accomplish this, we
recorded the high-resolution spectrum shown in Fig.~\ref{fig5}, which resulted from nearly 14 days of counting time with the
Si detector described in Sec.~\ref{sec:Rad decay}. 

The efficiency of the Si detector had already been thoroughly calibrated at 12 energies between 12 and 81 keV, and fitted to a standard
efficiency function \cite{Wa06}.  Its efficiency is nearly constant between 23.9 and 25.2 keV, with $\epsilon_{25.2}$/$\epsilon_{23.9}$(Si)
= 0.982(2), where the uncertainty has been conservatively estimated considering the small difference in efficiencies.  With the extended calibration
described in Sec.~\ref{sec:Cal}, the efficiency of our HPGe detector now covers this same energy region.  For the HPGe detector,
the efficiency also changes very little over the short range from 23.9 to 25.2 keV, but its slope has the opposite sign: in this
case $\epsilon_{25.2}$/$\epsilon_{23.9}$(HPGe) = 1.015(2).  Together, these two efficiency ratios allowed us to extract the contribution
of the 23.9-keV $\gamma$ ray from the area of the $K$ x rays in the HPGe spectrum of Fig.~\ref{fig4} (left panel) based on the
Si-detector spectrum of Fig.~\ref{fig5}.  The total counts in the tin $K$ x-ray peaks, which resulted from this analysis, appears
in the top line of Table~\ref{table3}.

\subsection{\label{sec:Effrat} Efficiency ratios}

As seen from Eq.~\ref{alpha}, in order to determine $\alpha_K$ for the 65.7-keV transition from $^{119m}$Sn, we
require the efficiency ratio, $\epsilon_{\gamma\,65.7}$/$\epsilon_{K25.8}$.  We obtain this ratio from the following
relation:
\begin{equation}
\frac{\epsilon_{\gamma\,65.7}}{\epsilon_{K25.8}} = \frac{\epsilon_{\gamma\,88.0}}{\epsilon_{K22.6}} \cdot
 \frac{\epsilon_{\gamma\,65.7}}{\epsilon_{\gamma\,88.0}} \cdot \frac{\epsilon_{K22.6}}{\epsilon_{K25.8}}.
\label{effratio}
\end{equation}
We take $\epsilon_{\gamma\,88.0}$/$\epsilon_{K22.6}$ from  the $^{109}$Cd calibration measurement described in
Sec.~\ref{sec:Cal}, where the value itself appears in Table~\ref{table1}.  The ratio $\epsilon_{\gamma\,65.7}$/$\epsilon_{\gamma\,88.0}$
is determined from our well-established detector efficiency curve obtained via CYLTRAN Monte Carlo calculations (see Sec.~\ref{sec:overview}
and Ref.\,\cite{He03}), while $\epsilon_{K22.6}$/$\epsilon_{K25.8}$ comes from the extension to that curve based on our new result from
$^{109}$Cd.  Both efficiency-curve ratios are over rather small energy differences, 22.3 keV in one case and 3.2 keV in the other, and both
represent efficiency changes of only a few percent.

The values for all four efficiency ratios from Eq.~\ref{effratio} appear in Table~\ref{table3}.

\subsection{\label{sec:Att} Attenuation in the sample}

As described in Sec.\,\ref{sec:sourceprep}, our source was obtained from an activated tin foil 6.8$\mu$m thick,
which, despite some deterioration during the lengthy irradiation process, yielded a large portion that remained
flat and could be used for our measurement.   We obtained the attenuation both of the tin x rays and of the
65.7-keV $\gamma$ ray using standard tables of attenuation coefficients \cite{Ch05}.  Since we are aiming at the
evaluation of $\alpha_K$ from Eq.~\ref{alpha}, what is important in that context is the attenuation for the x rays
relative to that for the $\gamma$ ray.  We determined that the x rays suffered 1.4(1)\% more attenuation than the
$\gamma$ ray and it is this result which appears as ``Relative attenuation" in Table\,\ref{table3}.

\begin{table}[t]
\caption{\label{table3}Corrections to the $^{119m}$Sn $K$ x rays and the 65.7-keV $\gamma$ ray as well as the additional
information required to extract a value for $\alpha_K$. }
\vspace{2mm}
\begin{ruledtabular}
\begin{tabular}{lll}
Quantity   &  Value  & Source  \\
\hline \\[-2mm]
\multicolumn{3}{l}{Sn ($K_{\alpha} + K_{\beta}$) x rays}  \\
~~ Total counts & 1.763(4) $\times 10^7$ & Sec.~\ref{xray} \\
~~ Impurities  & -1.46(3)$\times 10^5$  & Sec.~\ref{sec:Imp ident}  \\
~~ Lorentzian correction  &  +0.12(2)\%  &  Sec.~\ref{sec:Lor} \\
~~ Net corrected counts, $N_{K25.7}$  & 1.750(4)$\times 10^7$  &  \\
\hline \\[-2mm]
\multicolumn{3}{l}{$^{119}$Sn 65.7-keV $\gamma$ ray}  \\
~~ Total counts & 1.429(16)$\times 10^4$ &  \\
~~ Impurities  &  -1.010(13)$\times 10^3$  &  Sec.~\ref{sec:Imp ident} \\
~~ Net corrected counts, $N_{\gamma\,65.7}$ & 1.328(16)$\times 10^4$ &  \\
\hline \\[-2mm]
\multicolumn{3}{l}{Efficiency calculation}  \\
~~ $\epsilon_{\gamma\,88.0}$/$\epsilon_{K22.6}$ & 1.069(8) & Table~\ref{table1} \\
~~ $\epsilon_{\gamma\,65.7}$/$\epsilon_{\gamma\,88.0}$ & 1.0199(15) & \cite{He03} \\
~~ $\epsilon_{K22.6}$/$\epsilon_{K25.8}$ & 0.9568(20) & \cite{He03} \\
~~ $\epsilon_{\gamma\,65.7}$/$\epsilon_{K25.8}$ & 1.043(8) & \\
\hline \\[-2mm]
\multicolumn{3}{l}{Evaluation of $\alpha_K$}  \\
~~ $N_{K25.7}/N_{\gamma\,65.7}$  &  1318(16)  & This table  \\
~~ Relative attenuation  &  +1.4(1)\%  &  Sec.\,\ref{sec:Att} \\
~~ $\omega_K$  &  0.860(4)  &  \cite{Sc96}  \\
~~ $\alpha_K$ for 65.7-keV transition  & 1621(25)   &  Eq.\,\ref{alpha} \\
\vspace{-10.pt}
\end{tabular}
\end{ruledtabular}
\end{table}

\subsection{\label{sec:Lor} Lorentzian correction}

To be consistent with our previous efficiency-calibration procedures, we extracted our experimental peak areas
using the GF2 program (see Sec.\,\ref{sec:analysis}).  Specifically, we use a special modification of this
program that allows us to sum the total counts above background within selected energy limits.  To correct
for possible missed counts outside those limits, the program adds an extrapolated Gaussian tail.  We have noted
in previous papers \cite{Ni04,Ni05,Ni07,Ni08,Ni09} that this extrapolated tail does not do full justice to x-ray
peaks, whose shapes reflect the finite widths of the atomic levels responsible for them.  To correct for this
effect we computed simulated spectra using realistic Voigt-function shapes for the x-ray peaks and analyzed them
with GF2 following exactly the same fitting procedure as was used for the real data to ascertain how much was missed
by this approach.

The resultant correction factor appears in Table\,\ref{table3}.

\begin{table}[b]
\caption{\label{table4}Comparison of the measured $\alpha_K$ values for the 65.660(10)-keV $M$4 transition from $^{119m}$Sn
with calculated values based on different theoretical models for dealing with the $K$-shell vacancy.  Shown also
are the percentage deviations, $\Delta$, from the experimental value calculated as (experiment-theory)/theory.  For
a description of the various models used to determine the conversion coefficients, see text and Ref.\,\cite{Ni04}.}
\vspace{2mm}
\begin{ruledtabular}
\begin{tabular}{lll}
\multicolumn{1}{l}{Model}  & \multicolumn{1}{c}{~~$\alpha_K$} & \multicolumn{1}{c}{~~$\Delta$(\%)}  \\
\hline \\[-3mm]
Experiment & 1621(25)  &   \\
Theory: & & \\
~~~No vacancy  & 1544(1) & +5.0(16)  \\
~~~Vacancy, frozen orbitals  & 1618(1) & +0.2(16)  \\
~~~Vacancy, SCF of ion  & 1603(1) & +1.1(16)  \\
\vspace{-10.pt}
\end{tabular}
\end{ruledtabular}
\end{table}

\section{\label{sec:randd} Results and discussion}

The various correction terms and the results of our analysis are given in Table~\ref{table3}, where our final value
for the $K$-conversion coefficient itself, $\alpha_K$ = 1621(25), appears on the bottom line.  Its relative precision,
$\pm$1.5\%, is dominated by the $\pm$1.1\% uncertainty due to the counting statistics associated with the weak 65.7-keV peak.

There have been two previously reported measurements of this $K$ conversion coefficient, 1860(150) \cite{Dr71} and 1610(82) \cite{Ab75}.
Both were published more than 35 years ago and both were quoted with considerably larger uncertainties than ours.  
Our result is outside the 1$\sigma$ error bars of the earlier of the two measurement but is consistent with the more
recent one.

Our measured $\alpha_K$ value is compared with three different theoretical calculations in Table\,\ref{table4}.  
All three calculations have been made within the Dirac-Fock framework, but one ignores the presence of the $K$-shell
vacancy while the other two include it using different approximations: the frozen orbital approximation, in which
it is assumed that the atomic orbitals have no time to rearrange after the electron's removal; and the SCF
approximation, in which the final-state continuum wave function is calculated in the self-consistent field
(SCF) of the ion, assuming full relaxation of the ion orbitals.  To obtain these results we used the value
65.660(10) keV \cite{Sy09} for the $^{119m}$Sn transition energy.  The experimental uncertainty in this
number is reflected in the uncertainties quoted on the theoretical values of $\alpha_K$ in the table.

\begin{figure}[t]
\epsfig{file=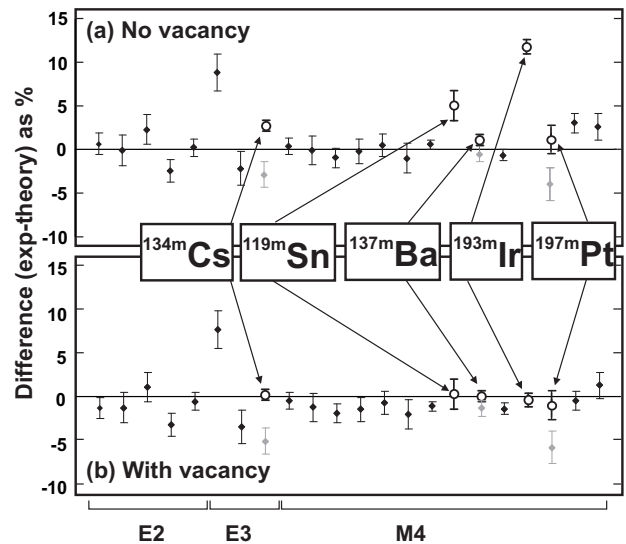,width=8.1cm}
\caption{Percentage differences between the measured and calculated ICCs for two Dirac-Fock calculations: one (a)
is without the atomic vacancy and the other (b) is with it included in the ``frozen orbital" approximation.  The points
shown as solid diamonds in both plots correspond to the twenty cases listed by Raman {\it et.\,al} \cite{Ra02} with better
than 2\% precision; as indicated at the bottom, five are for $E$2 transitions, three for $E$3 and the remainder are for
$M$4 transitions.  The points shown as open circles correspond to our five more-recently measured $\alpha_K$ values.  For
the cases of $^{134m}$Cs \cite{Ni07,Ni08}, $^{137m}$Ba \cite{Ni07,Ni08} and $^{197m}$Pt \cite{Ni09}, the earlier Raman
values are shown in grey: for $^{193m}$Ir \cite{Ni04,Ni05} and for $^{119m}$Sn, the case presented here, there were no
earlier values with sub-2\% precision.}
\label{fig6}
\end{figure}

The percentage deviations given in Table\,\ref{table4} indicate acceptable agreement between our measured
result and the two calculations that include some provision for the atomic vacancy.  Our measurement disagrees
by 3.1 standard deviations from the calculation that ignores the vacancy.  That point is convincingly illustrated
by Fig.\,\ref{fig6}, in which our measurement (open circles labeled $^{119m}$Sn) is compared graphically with
the no-vacancy and the ``frozen orbital" vacancy calculations.  We have now made five precise ICC measurements, 
which together present a consistent pattern that supports the inclusion of the atomic vacancy when calculating
conversion coefficients.

\section{\label{sec:theory} Theoretical calculations of vacancy-effect systematics} 

As a guide to future experiments, we look here at the influence of the $K$-shell conversion vacancy on ICC values as a
function of atomic number, conversion-electron energy and transition multipolarity. 

The ICC calculations were performed by the Dirac-Fock method using approximations and expressions described in detail in
Refs.\,\cite{Ra02,Ba02,Ni04}. Note that advantage was taken of the surface-current model for the uniform distribution of a charge
over the volume of a spherical nucleus with radius $R_0=1.2A^{1/3}$~fm, where $A$ is the mass number. We used the
experimental binding energy of the $K$ electron, $\varepsilon_K$, to determine the conversion-electron energy using the
following expression:
\begin{equation}
E_k=E_{\gamma}-\varepsilon_K,
\end{equation}
where $E_{\gamma}$ is the $\gamma$-ray energy.

\begin{figure}[t]
\epsfig{file=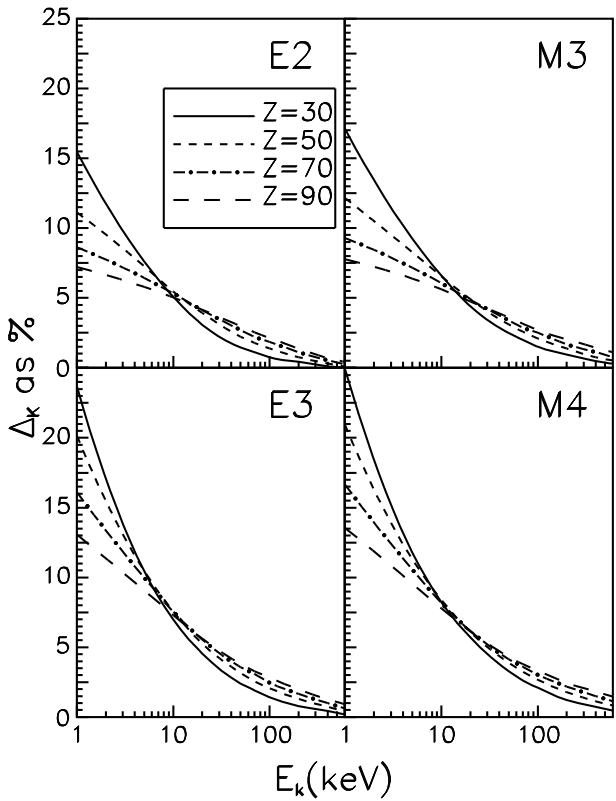,width=8.1cm}
\caption{The difference $\Delta_K$ between $\alpha_K(\rm v)$ and $\alpha_K(\rm nv)$ expressed as a percent (see Eq.~\ref{delta}) is
plotted versus the conversion-electron energy $E_k$ for elements with $Z=$~30, 50, 70, and 90.  Four multipolarities, $E$2, $E$3, $M$3, and $M$4,
are considered.}
\label{fig7}
\end{figure}

We calculated each ICC in two ways: one which used the framework of the "frozen core" model to account for the vacancy after
conversion, denoted $\alpha_K(\rm v)$; and one which ignored the vacancy, denoted $\alpha_K(\rm nv)$.  In
Fig.~\ref{fig7} we plot the differences, $\Delta_K$, between the two calculated results, where
\begin{equation}
\Delta_K=\frac{\alpha_K(\rm
v)-\alpha_K(\rm nv)} {\alpha_K(\rm v)}\times 100 \%.
\label{delta}
\end{equation}
These differences are plotted as a function of $E_k$ for four representative elements with $Z=$~ 30, 50, 70, and 90 and for
several different transition multipolarities.

As is evident from Fig.~\ref{fig7}, the difference $\Delta_K$ depends strongly on the conversion-electron energy, especially at
low values of $E_k$ and for high-multipolarity transitions. For example, in the case of an $M$4 transition at $E_k=$~1~keV, the
difference is as large as $25 \%$ for $Z=$~30 and $13.5 \%$ for $Z=$ 90.  Furthermore, at low $E_k$ values $\Delta_K$ is not only large
but it also differs very considerably from one $Z$ value to another.  As $E_k$ increases, however, $\Delta_K$ decreases, so that
above 10~keV it drops below $\sim$8\% and the four curves for the different $Z$ values are rather close to one another. At high
energies, $E_k \approx$~500-600 keV, $\Delta_K$ still persists at the level of about 1$\%$ for the heavier elements. It is interesting
to note that at low energies, $E_k \la 10$~keV, the $\Delta_K$ decreases with atomic number $Z$, whereas at $E_k \ga 10$~keV $\Delta_K$
increases with $Z$ for all cases shown.

The difference $\Delta_K$ is shown to increase with the transition multipolarity at all energies $E_k$, but especially at low $E_k$
values. This is attributable to the selection rules, by which the high multipolarity ICCs involve final continuum electron wave
functions with large orbital momenta $\ell$: For example, $g_{7/2}$ and $g_{9/2}$ wave functions contribute to $\alpha_K^{E4}$, and
$f_{7/2}$ and $h_{9/2}$ ones contribute to $\alpha_K^{M4}$ \cite{Tr10}.  Because of this, the associated conversion matrix elements
turn out to be more sensitive to the presence of the vacancy.

In addition, one can see that the differences $\Delta_K$ for $E$L and $M$(L+1) transitions are similar to one another (e.g. $E$2 and $M$3,
$E$3 and $M$4 in Fig.~\ref{fig7}). Here the important point is that matrix elements for $E$L and $M$(L+1) transitions in the case of
conversion in the $K$ shell involve continuum wave functions with the same $\ell$ values. For example, for an $E$2 transition the $d_{3/2}$
and $d_{5/2}$ states are the final-state wave functions for $K$-shell conversion, with $d_{3/2}$ being the largest one; while for an $M$3 transition
the $d_{5/2}$ and $f_{7/2}$ final states are involved, with $d_{5/2}$ being the largest one.  Because radial matrix elements involving
$d_{3/2}$ and $d_{5/2}$ functions are closely allied, the relative magnitudes $\Delta_K$ are approximately the same for $E$L and
$M$(L+1) transitions.

Evidently, while the effect of the vacancy -- as expressed by $\Delta_K$ -- shows some dependence on $Z$, the dependence on $E_k$ is much
more pronounced. It is therefore interesting to note the conversion-electron energies for the four precise measurements that we have
previously reported \cite{Ni04,Ni05,Ni07,Ni08,Ni09}: 4 keV ($^{193m}$Ir), 92 keV ($^{134m}$Cs), 268 keV ($^{197m}$Pt) and 624 keV ($^{137m}$Ba).
These energies can be compared with $E_k$=36 keV, the energy of the conversion electrons emitted in the decay of $^{119m}$Sn, the
case we report here.  Our new result fills in a region of energy not previously covered.

\section{\label{sec:conc} Conclusions}

Our measurement of the $K$-shell internal conversion coefficient, $\alpha_K$, for the 65.7-keV $M$4 transition from
$^{119m}$Sn has yielded a value that only agrees with versions of Dirac-Fock theory that include the atomic vacancy.
This result confirms the conclusion reached from our previous precise ICC measurements, and extends its validity down
to $Z$ = 50, a lower atomic number than we have studied previously. 
 
A look at Fig.\,\ref{fig6} will give the reader an appreciation of the current situation.  Several years ago, our
measurement of the 80.2-keV $M4$ transition in $^{193}_{~77}$Ir \cite{Ni04,Ni05} was the first to show definitively that the
atomic vacancy must be included in the theory, at least in the case of low-energy ($\sim$4 keV) $K$-conversion electrons.  
Our later measurements of transitions in $^{134}_{~55}$Cs and $^{137}_{~56}$Ba  \cite{Ni07,Ni08} represented the
first test among lighter nuclei and for transitions with significantly higher-energy conversion electrons.  They confirmed
our earlier conclusions by showing a clear preference for the Dirac-Fock theory that included provision for the atomic
vacancy; and they also removed an apparent anomaly for the case of $^{134m}$Cs by replacing earlier faulty experimental
results.  In the case of $^{197m}_{~~~78}$Pt we also corrected an earlier result to bring agreement with theory.

Our present result extends the test of ICC calculations down to $Z$ = 50 and at a conversion-electron energy of 36 keV.
It reaches the same conclusion.  With this result, it is becoming increasingly clear that Dirac-Fock calculations with the
atomic vacancy included provide a reliable way to determine ICCs to a precision no worse than $\pm$1\%.\\

\begin{acknowledgments}

We wish to thank Dr. Latha Vasudevan and the staff of the Texas A\&M Nuclear Science Center for their help in conducting neutron
activations.  The work of the Texas A\&M authors is supported by the U.S. Department of Energy under Grant No.\,DE-FG03-93ER40773
and by the Robert A. Welch Foundation under Grant no.\,A-1397.

\end{acknowledgments}

\end{document}